\documentclass{article}
\usepackage{nips2003e,times}
\usepackage{epsfig,graphics,subfigure}
\newcommand{\bc}{\begin{center}}
\newcommand{\ec}{\end{center}}
\newcommand{\bsc}{\begin{slide}\begin{center}}
\newcommand{\esc}{\end{center}\end{slide}}
\newcommand{\bt}{\begin{tabular}}
\newcommand{\et}{\end{tabular}}
\newcommand{\bfr}{\begin{flushright}}
\newcommand{\efr}{\end{flushright}}
\newcommand{\bfl}{\begin{flushleft}}
\newcommand{\efl}{\end{flushleft}}

\title{Message passing in random satisfiability problems}
\author{Marc M\'ezard\\
CNRS, Laboratoire de Physique Th\'eorique et Mod\`eles Statistiques\\
Universit\'e Paris Sud\\
B\^atiment 100, 91405 Orsay Cedex, France\\
\texttt{mezard@lptms.u-psud.fr}
}

\begin{document}

\maketitle

\begin{abstract}
This talk surveys the recent development of message passing procedures
for solving constraint satisfaction problems. The cavity method from
statistical physics provides a generalization of the belief propagation
strategy that is able to deal with the clustering of solutions in these
problems. It allows to derive  analytic results on their phase diagrams, 
and offers
a new algorithmic framework.
\end{abstract}

\section{ Message passing}
Many NIPS participants can be confronted to the following type of generic problem
\cite{MM_scpersp}.
A system involves many 'simple'  discrete variables, and each variable can take a certain number
of discrete values (which is not too large). These variables interact, and the interactions 
can be written in the form of
constraints, where each constraint  involves a small fraction of all the variables.
One wants to find 
 the values of the variables which are  compatible with all constraints (or, when this is impossible,
find the values  which minimize the number of violated constraints).

 This 'constraint satisfaction problem' (CSP) is a ubiquitous
situation which shows up for instance in statistical physics,
combinatorial optimization, error correcting codes, statistical
inference,... It turns out that, in order to solve CSPs, these various disciplines have
developed, often independently, some message passing procedures which
turn out to be remarkably powerful.  The general
strategy involves two main steps:
\noindent
\begin{itemize}
\item Organize the constraints and variables in a graph
\item Exchange messages, of probabilistic nature, along the graph
\end{itemize}
We shall illustrate it here using as an example the famous 'satisfiability' problem.

\section{Satisfiability and complexity theory}
The problem of satisfiability involves $N$ Boolean variables ${ x_i
 \in \{0,1\} }$. There exist thus { $2^N$} possible configurations of
 these variables. The constraints take the special form of 'clauses',
 which are logical 'OR' functions of the variables. For instance the
 clause $ x_1 \vee x_{27} \vee \bar x_3$ is satisfied whenever $x_1=1$
 or $x_{27}=1$ or $x_3=0$. Therefore, among the $8$ possible
 configurations of $ x_1, x_{27}, x_3$, the only one which is
 forbidden by this clause is $x_1=x_{27}=0$, $x_3=1$. An instance of
 the satisfiability problem is given by the list of all the clauses.

Satisfiability is a decision problem. One wants to give a yes/no
answer to the question: is there a choice of the Boolean variables (called an 'assignment')
such that all constraints are satisfied (when there exists such a
choice the corresponding instance is said to be 'SAT', otherwise it is
'UNSAT')?

This problem appears in many fields and its importance is a
consequence of its genericity: an instance of the SAT problem can be
seen as a logical formula (such as {$(x_1 \vee x_{27} \vee \bar
x_3) \wedge (\bar x_{11} \vee x_2) \wedge \ldots$}), and it turns out
that any formula can be written in this form of the 'and' function of
several clauses, called 'conjunctive normal form'.

Satisfiability plays an essential role in the theory of computational
complexity,because it was the first problem which has been shown to be
'NP-complete', in a beautiful theorem by Cook in 1971 \cite{Cook}.  The
NP problems are those decision problems where a 'yes' answer can be
checked in a time which is polynomial in $N$. This is a vast class of
problems which contains such difficult problems as the traveling
salesman, the problem of finding a Hamiltonian path in a graph (a path
which goes once through all vertices), the protein folding problem, or
the spin glass problem in statistical physics. The fact that
satisfiability is NP-complete means that any other NP problem can be
mapped to satisfiability in polynomial time. So if we happened to have
a polynomial algorithm for satisfiability (i.e. an algorithm that
solves the decision problem in a time which grows like a power of
$N$), all the problems in NP would also be solved in polynomial
time. This is generally considered unlikely, but the corresponding
mathematical problem (whether the NP class is distinct or not from the
'P' class of problems which are solvable in polynomial time) is an
important open problem.

The result of Cook is a worst case analysis of the
satisfiability problem. However it is known experimentally that many
instances of this problem are easy to solve, and researchers have
started to study some classes of instances in order to characterize
the {\bf{'typical case'}} complexity of satisfiability problems. A
class of instances is a probability measure on the space of
instances. A much studied problem in this framework is the random
'3-SAT' problem. Each clause contains exactly three variables; they
are  chosen randomly in $\{ x_1,..,x_N\}$ with uniform
measure, and each variable is negated randomly with probability
$1/2$. This problem is particularly interesting because its difficulty
can be tuned by varying one single control parameter, the ratio $
\alpha=\frac{M}{N}$ of constraints per variable. One expects
intuitively that for small $\alpha$ most instances are SAT, while for
large $\alpha$ most of them are UNSAT. Numerical experiments have
confirmed this scenario, but they indicate actually a more interesting
behavior. The probability that an instance is SAT exhibits a sharp
crossover, from a value close to $1$ to a value close to 0, at a
threshold $\alpha_c$ which is around $4.3$. When the number of
variables $N$ increases, the crossover becomes sharper and sharper
\cite{KirkSel,Crawford}. It has been shown that it becomes a staircase behavior at
large $N$ \cite{Friedgut}: almost all instances are SAT for
$\alpha<\alpha_c$, almost all instances are UNSAT for
$\alpha>\alpha_c$, and some bounds on the value of $\alpha_c$ have
been derived\cite{bounds_4,bounds_5}. This threshold behavior is nothing but a phase
transition as one finds in physics, and has been analyzed using the
methods of statistical physics \cite{KirkSel,MZKST}.

A very interesting observation is that the algorithmic difficulty of
the problem, measured by the time taken by the algorithm to answer if
a typical instance is satisfiable, also depends strongly on $\alpha$:
the typical problem is easy when $\alpha $ is well below or well above
$\alpha_c$, and is much harder when $\alpha$ is close to
$\alpha_c$. Therefore the region of phase transition is also the
region which is interesting from the computational point of view.


\section{
Statistical physics of the  random 3-SAT problem}

In the last two decades, it has been realized that some concepts and
methods developed in statistical physics can be useful to study
optimization problems \cite{TCS_issue}. The interest in this type of approach
has been focusing in recent years on the random satisfiability problem,
following the works of colleagues like Kirkpatrick, Monasson and
Zecchina \cite{MZKST,MoZe_pre,BMW}. This is indeed a choice problem for a
multidisciplinary study, because the SAT-UNSAT transition is a phase
transition in the usual statistical physics sense, and the algorithmic
complexity of the problem shows up precisely in the neighborhood of
this phase transition. On the analytical side, the first
breakthrough used the replica method and computed some approximations
of the phase diagram using either a 'replica symmetric' approximation,
or some variational approximation to the replica symmetry breaking
solution. Hereafter I will mainly review  the most recent
developments initiated in \cite{MEPAZE,MZ_pre}. These have provided some analytical insight
into the phase diagram of the problem, which turns out to be more
complicated than originally thought, as well as a new algorithmic
strategy based on message passing for the case of random
3-satisfiability.  

\begin{figure}[ht]
\begin{center}
\epsfig{file=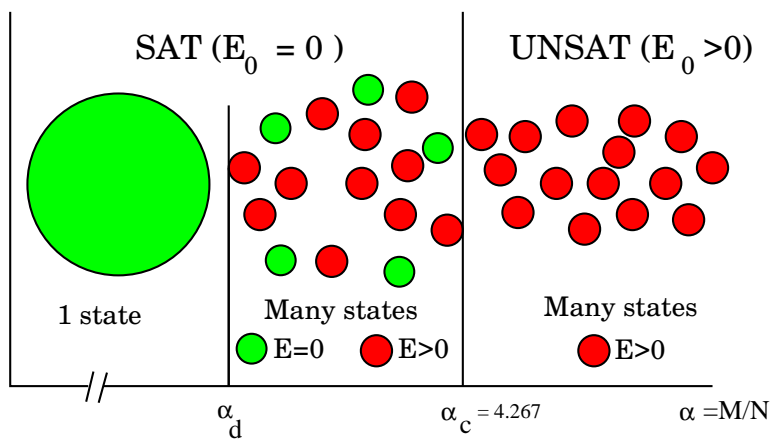,width=0.8\textwidth}
\end{center}
\caption{The phase diagram of the random 3-SAT problem: when one increases
the ratio $\alpha$ of constraints to variables, one finds successively
an Easy-SAT phase, a Hard-SAT phase, and an UNSAT phase}
\label{phase_diag}
\end{figure}

The main analytical result on the phase diagram is summarized in
Fig.~\ref{phase_diag}. When one varies the control parameter $\alpha$ (the
number of constraints per variable), there actually exist three
distinct phases, separated by two thresholds $\alpha_d$ and
$\alpha_c$.  The threshold at $ \alpha_c =4.267$ is the satisfiability
threshold which separates the SAT phases at $\alpha<\alpha_c$ from the
UNSAT phase at $\alpha>\alpha_c$.  But in the SAT region,there
actually exist two distinct phases. They differ by the structure of
the space of SAT assignments (the 'SAT space').  For
$\alpha<\alpha_d\simeq 3.86$, the set of SAT assignments builds one cluster
which is basically connected \cite{BMW,Par_remarks}. Starting from a generic SAT assignment
(which is a point on the unit N-dimensional hypercube), one can step
to another one nearby by flipping a finite number of variables. It
turns out that one can get from any
generic SAT assignment to any other one through a sequence of such steps,
always staying in the SAT space. On the other hand, for
$\alpha>\alpha_d$, this SAT space becomes disconnected: the SAT
assignments are grouped into many clusters. It is impossible to get
from one cluster to another one without having to flip at some moment
a macroscopic fraction of the variables. Simultaneously, the space of
configurations develops many 'metastable states': imagine walking
along the configuration space (the N-dimensional hypercube), flipping
one variable at a time, but allowing only the moves which decrease, or
leave constant, the number of violated clauses. Such an algorithm will
get trapped into some 'metastable clusters', which are connected
clusters of assignments, all violating the same (non-zero) number of
clauses.  This structure has some direct algorithmic implication: the
separation of clusters in the phase $\alpha>\alpha_d$ makes it very
difficult for algorithms based on local moves to find a SAT
assignment. The phase $\alpha<\alpha_d$ is thus called the 'Easy-SAT'
phase, while the one between $\alpha_d$ and $\alpha_c$ is the
'Hard-SAT' phase.

These results have been obtained using the cavity method. Originally
this method was invented in the study of spin glass theory, in order
to solve the Sherrington Kirkpatrick model \cite{MPV}. It is only
recently that new developments in this method allowed to solve 'finite
connectivity' problems where each variable interacts with a small
number of other variables\cite{MP_Bethe}. Thanks to these development, it
has become possible to study constraint satisfaction
problems. Although the method is not fully rigorous, the
self-consistency of the main underlying hypothesis has been checked
\cite{MonRic,MertMZ,MonParRic}, and therefore the results concerning the satisfiability
threshold, as well as crucial features of the phase diagram like the
existence of the intermediate Hard-SAT phase, are
conjectured to be exact results, not approximations. 
Of course, it is very important to 
develop rigorous mathematical approaches in order to confirm or infirm them.

It turns out that the cavity method when applied to one given instance of
the problem amounts to a message passing procedure which can also be used as an
algorithm. Therefore one surprising offspring of the theoretical development
of the cavity method in recent years, which aimed at answering
very fundamental questions concerning the phase diagram, has been the finding
of a new class of algorithms, which turn out to be very efficient in the
Easy-SAT phase.

The next sections aim at describing the main ideas of the cavity
method, taking the point of view of the message passing algorithm. It
is impossible to explain all the details, or to give the
justification, for all these steps. The interested reader is referred
to refs \cite{MP_Bethe} for the basic concepts of the cavity method in
finite connectivity disordered systems, and to refs 
\cite{MEPAZE,MZ_pre,BMZ,MertMZ} for the
solution of the random satisfiability problem.

\section{Factor graphs and warning propagation}
\begin{figure}[ht]
\begin{center}
\epsfig{file=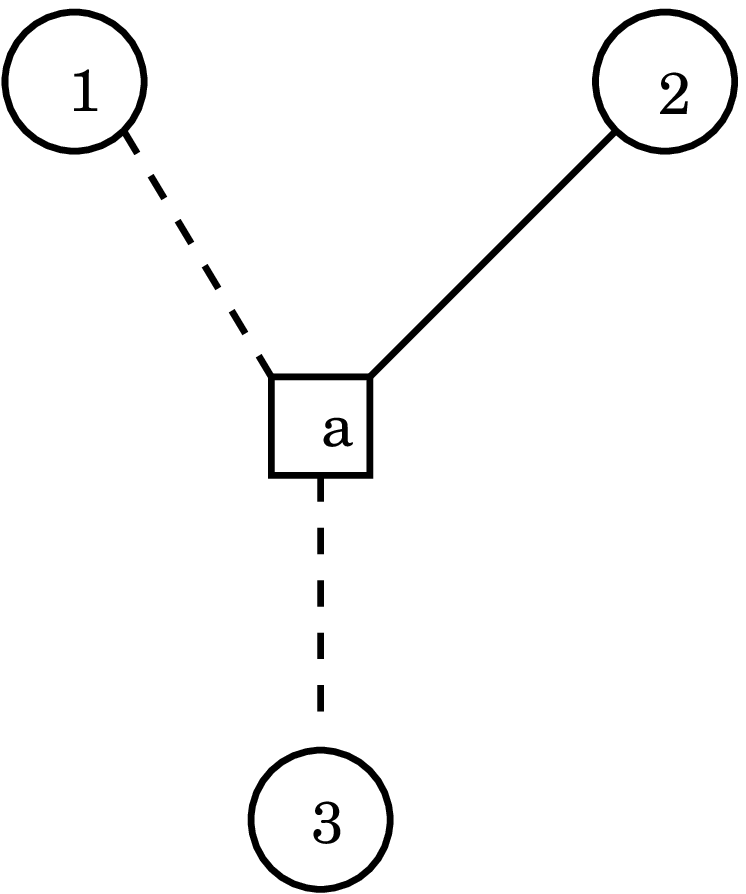,width=0.2\textwidth}
\hspace {3 cm}
\epsfig{file=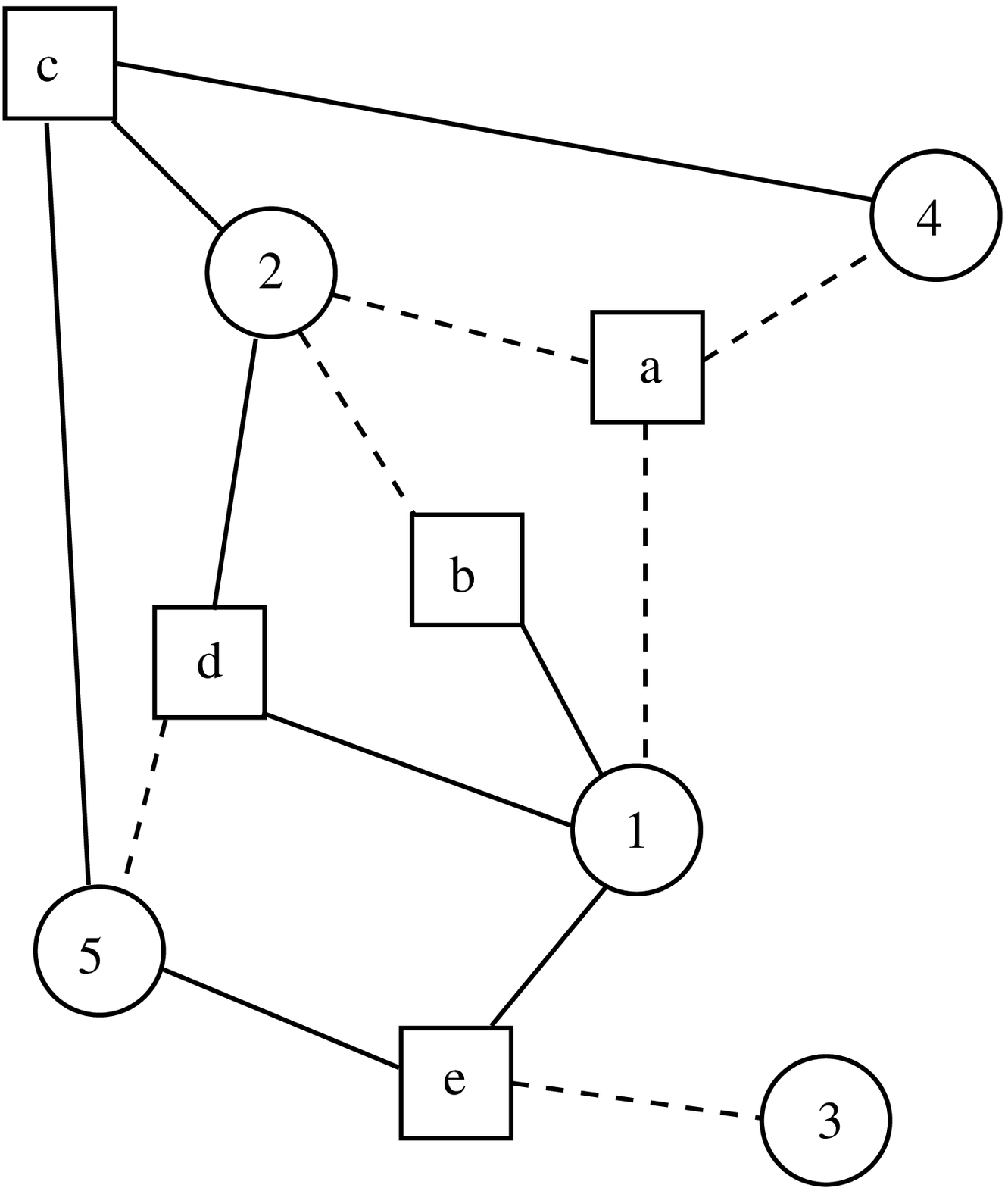,width=0.35\textwidth}
\end{center}
\caption{Factor graph representation of satisfiability: A variable is represented by
a circle. A clause is represented by a square, connected with a full (resp. dashed)
 line to a variable when this variable appears as such (resp. negated) in the clause.
Left hand side: The clause $ \bar x_1 \vee x_2 \vee \bar x_3$. Right hand side:
the factor graph representing the formula: 
${( \bar x_1 \vee  \bar x_{2} \vee \bar x_4) }{\wedge }{ ( x_{1} \vee 
\bar x_2)}{\wedge }{( x_2 \vee   x_{4} \vee  x_5) } {\wedge }{ ( x_{1} \vee 
 x_2 \vee \bar x_5)}{\wedge }{(  x_1 \vee \bar  x_{3} \vee  x_5) }$}
\label{fig_clause}
\end{figure}

The first step consists in a graphical representation of the satisfiability
problem. Each clause
is represented by a function node, connected to the various variables which
appear in the clause, as described in Fig.\ref{fig_clause}.

Factor graphs turn out to be very useful in various contexts, including  statistical 
inference  \cite{pearl} and error correcting codes   \cite{codes_special_issue}. 
The general formalism is described in \cite{factor_graphs}.

In a random 3-SAT problem with $N$ variables and $M=\alpha N$ clauses,
in the large $N$ limit, it is easy to see that the factor graph is a
random bipartite graph, where the function nodes have a fixed
connectivity equal to 3, and the variable nodes have a Poisson
distributed connectivity with a mean $3 \alpha$.

The simplest case of message passing is the warning propagation.
Along all the edges of the graph, one passes  messages. If clause
$a$ involves variable $i$, the message  $u_{a \to i} \in\{ 0,1\}$ sent from clause
$a$ to variable $i$ is a way for clause $a$ to inform variable $i$:
A warning {  $u_{a \to i}=1$} means :
 ``According to the messages I have received (from the other variables which
are connected to me), you should take the value which satisfies me!''.
When $u_{a \to i} =0$, this means that no warning is passed, which means a message
saying:  ``According to the messages I have received, there is no problem, 
you can take any value!''
\begin{figure}[ht]
\begin{center}
\epsfig{file=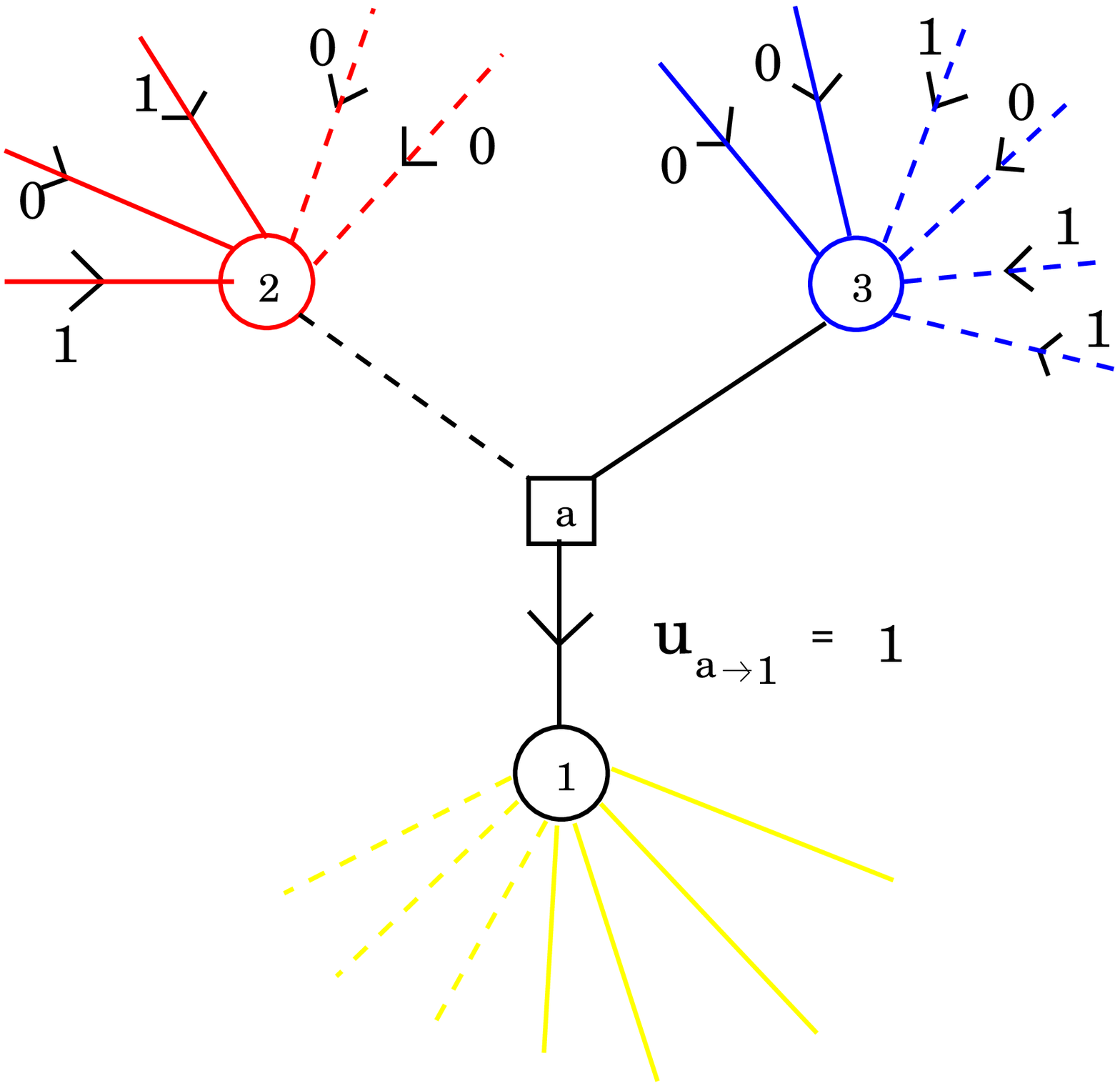,width=.4\textwidth}
\hspace{2 cm}
\epsfig{file=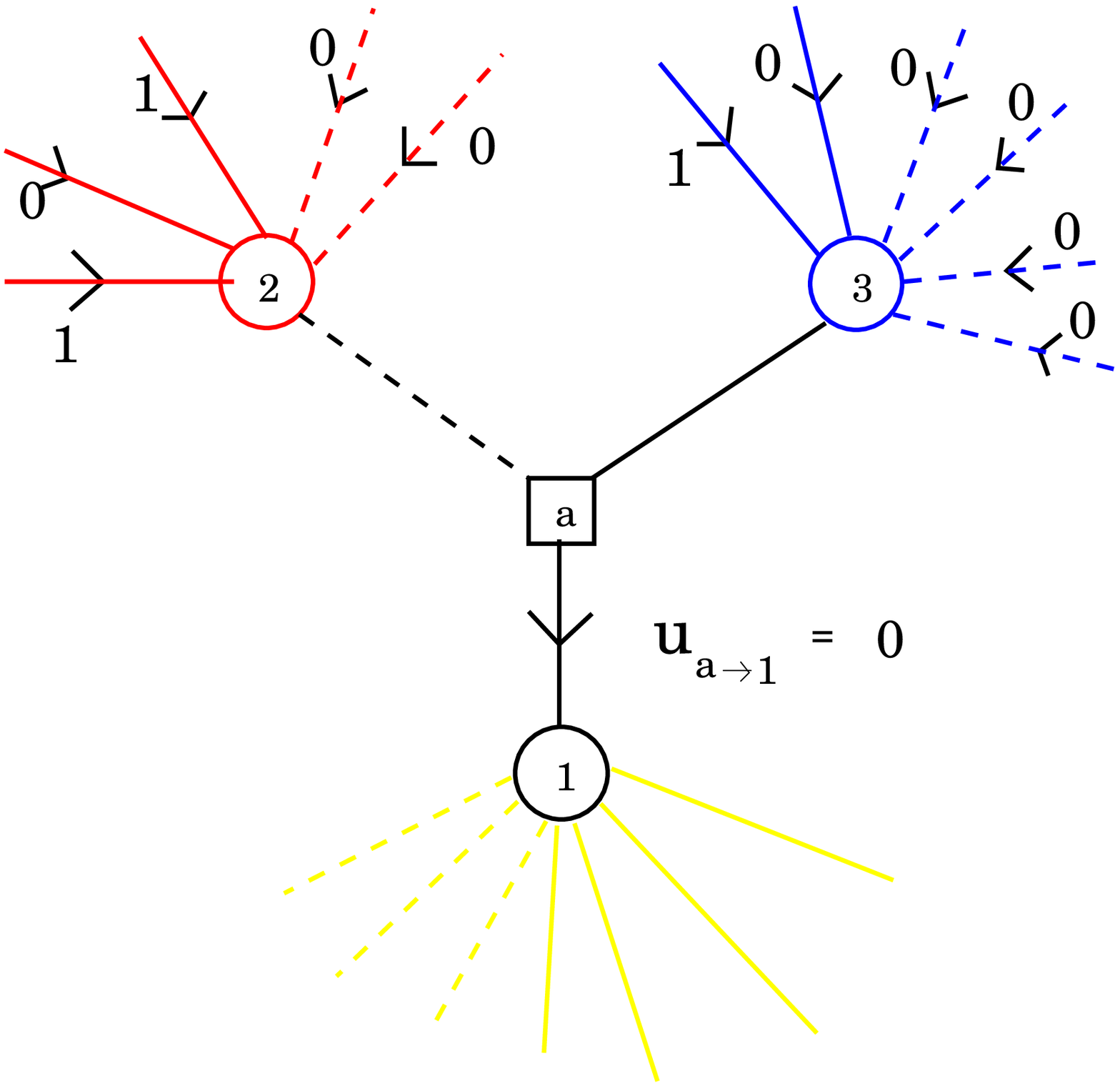,width=.4\textwidth}
\end{center}
\caption{Two examples of warning propagation in a clause $a$. In order to determine
the message which it passes to variable $1$, clause $a$ considers all the messages 
received by the other variables to which it is connected, here the variables $2$ and $3$.
On the left hand side, the messages received by $2$ tell it to take the value $x_2=1$, 
which does not satisfy clause $a$, and the messages  received by $3$ tell it to take 
the value $x_3=0$, which does not satisfy clause $a$. Therefore, in order to be satisfied, 
clause $a$ must rely on
variable $1$. It thus sends a warning $u_{a \to 1}=1$. Right hand side: an example
where no warning is sent, because variable $3$ is told by its environment to take the value
$x_3=1$, which satisfies clause $a$.}
\label{clause_prop}
\end{figure}

It can be shown that warning propagation converges and gives the correct 
answer in a satisfiability problem described by a tree factor graph (such a problem can
 be nontrivial if there are, among the clauses, some 'unit clauses', involving 
a single variable, and forcing this variable): the problem is  SAT if and only if, 
on each variable node, there are no contradictory messages.

This simple warning propagation corresponds to some limiting case of the
celebrated belief propagation (``BP'') algorithm. The basic idea of BP
is to study the probability space of all SAT assignments with uniform measure.
Consider again the clause in Fig.\ref{clause_prop}. Imagine one 
knows the joint 'cavity' probability $P^{(a)}(x_2,x_3)$ for the variables $x_2,x_3$
when the clause $a$ is absent. Then one can deduce the following estimate
 from $a$ concerning the state of $1$:
\begin{equation}
P_{a \to 1}(x_1)=  \sum_{x_2,x_3} 
C_a(x_1,x_2,x_3) P^{(a)}(x_2,x_3) \ , \label{bp1}
\end{equation}
where $C_a$ is the indicator function of clause $a$, equal to one if and only if the clause
is satisfied. The
 probability distribution of $x_1$ in the absence of a
clause $b$ is given by:
\begin{equation} 
P^{(b)}(x_1)= C \prod_{a \in V(1)\setminus b} P_{a \to 1}(x_1)
\label{bp2}
\end{equation}
where $C$ is a normalization constant.

These exact equations are useless as such, but they become very useful if one
adds the hypothesis that
\begin{equation}
 P^{(a)}(x_2,x_3) \sim P^{(a)}(x_2) P^{(a)}(x_3) \ .
\label{factorization}
\end{equation}
 Then eqs (\ref{bp1},\ref{bp2})
close onto a set of self-consistent equations for the cavity probabilities
$P^{(a)}(x)$. These equations can be interpreted as a message passing procedure
which is the BP algorithm; they also carry the name of Bethe approximation in
statistical physics. The warning propagation equations are some version of BP
in which one focuses onto the variables which are forced to take one given
value in all the SAT configurations (in the statistical physics 
approach they correspond to writing the Bethe equations directly at 
zero temperature). The onset of clustering at this zero temperature level
signals the existence of clusters in which a given variable is frozen (it
takes the same value in all configurations of the cluster): it takes place at a value $\alpha \sim 3.91$.

The main question is whether the factorization approximation (\ref{factorization})
is correct. If the factor graph is a tree, it is obviously correct.
In the case of a random 3-SAT, or random K-SAT problem, one may notice that generically the factor graph is locally tree-like: the loops appear only at
large distances (of order $\log N$). Therefore, in the absence of clause $a$,
the two variables $x_2,x_3$ are far apart. In such a situation one may hope
that the  factorization will hold, in the large $N$ limit.
However this is true only in the case where there is a single 
pure state in the system (i.e. a single cluster of solutions).
If there are several pure states, it is well known in statistical physics that the correlations of the variables decay at large distances only when the
probability measure is restricted to one pure state.

\section{
Proliferation of states: survey propagation } 
Therefore one can expects the BP, or the warning propagation, to be correct
in the Easy-SAT phase. In the Hard-SAT phase where pure states proliferate,
the BP would be correct if we had a way to restrict the measure to the 
SAT configurations in one given cluster. However there is no way to 
achieve this globally. BP is a local message passing procedure. Locally it will
tend to find equilibrated configurations corresponding to one cluster, but there is no way to select the same cluster in distant parts of the factor graph.

In order to handle such a situation the cavity method introduces
generalized messages: Along each edge $a-i$, a message is sent from
clause $a$ to variable $i$. This message is a survey of the elementary
messages in the various clusters of SAT configurations.  Because
warnings are so simple (this is where it is useful to use warnings
rather than standard BP), the survey is characterized by a single real
number $\eta_{a \to i}$, which gives the probability of a warning
being sent from constraint $a$ to variable $i$, when a cluster is
picked up at random in the set of all clusters of SAT assignments.


The survey propagation (SP) equations are easily written. Looking again at the
 clause of Fig.\ref{clause_prop}, let us denote by $U$ the set of
function nodes, distinct from $a$, connected through a full line to variable $x_2$, and 
$V$ the set connected through a dashed line. Introduce the probabilities
that no warnings are sent from $U$ and $V$, given by
$\pi^2_+=\prod_{b\in U} (1-\eta_{b \to 2})$ and 
$\pi^2_-=\prod_{b\in V} (1-\eta_{b \to 2})$. Similar quantities
$\pi^3_\pm$ are introduced for the variable $x_3$. Then the SP equations
read:
\begin{eqnarray}
\eta_{a \to 1} =
\frac {\pi^2_-(1-\pi^2_+)}{\pi^2_+ + \pi^2_- - \pi^2_+ \pi^2_-} \ \ 
\frac {\pi^3_+(1-\pi^3_-)}{\pi^3_+ + \pi^3_- - \pi^3_+ \pi^3_-}
\label{sp_eq}
\end{eqnarray}
It turns out that the propagation of these
surveys along the graph converges for a generic problem in the
whole Hard-SAT phase.

These SP equations can be used in two ways. When iterated on one given sample,
they provide very interesting information concerning each variable: From the
set of surveys, one can compute for instance the probability that one variable
is constrained to $0$, when one chooses a SAT cluster randomly. The survey
inspired decimation algorithm \cite{MEPAZE,MZ_pre} uses this information as
follows: it identifies the most biased variable (the one which is polarized to
1, or to 0, with the largest probability), and fixes it. Then the problem is
reduced to a new satisfiability problem with one variable less. One can run
again the SP algorithm on this smaller problem and iterate the procedure. This
simple strategy finds SAT assignment in the region $\alpha<4.252$ (quite
close to the UNSAT threshold $\alpha_c=4.267$), in a time of order $N \log N$
which allows to reach large sample sizes of a few million variables. Some
backtracking in the way one performs the decimation \cite{Par_backtr}
 allows to get closer to the  threshold, but until now it is not known if 
these survey based algorithms will
allow to solve the algorithmic problem in the whole Hard-SAT phase or not.

On the other hand one can also use the SP equations in order to find
analytical results. The idea is to perform a statistical analysis of the SP
equations: one introduces the probability density $P(\eta)$, when one picks up
an edge $a-i$ randomly in a large random 3-SAT problem, that the survey
$\eta_{a \to i}$ takes a given value $\eta$. The SP equations (\ref{sp_eq})
allow to write a self-consistent non-linear integral equation for $P(\eta)$.
Solving these equations, one can deduce all the features of the phase diagram
of Fig.\ref{phase_diag}. The thresholds found for random K-sat, 
for $K=3,4,5,6,10$, are respectively equal to
$4.267,\; 9.931,\; 21.12,\; 43.37,\; 708.9 $ (see \cite{MertMZ}).

The equations can also be generalized to study the UNSAT phase,
in which one can determine the minimal number of violated clauses, both 
analytically, and also algorithmically on one given sample,
but this goes beyond the present paper.

This general cavity method strategy can also be used in other constraint
satisfaction problems \cite{BMWZ}, and has found non-trivial results for
instance on the coloring problem\cite{colouring}. A particularly interesting
case is the XOR-SAT problem \cite{xorsat}, where the same type of phase diagram
is found, and the cavity method result can be checked versus rigorous
computations. It is also worth mentioning that for $K$ even, the SAT-UNSAT
threshold $\alpha_c$ found with the cavity method has been shown to be an
upper bound to the correct result\cite{FraLeo}. Our conjecture is that this bound is
actually tight.

 To summarize, we have considered sparse
 network of many interacting elements, with interactions taking
the form of constraints on their relative values. In this situation,
one often finds a ``Hard-SAT'' phase, with an
 exponential number of well separated solution clusters,
together with an exponentially larger number of 'metastable states'.
It turns out that the local
 exchange of probabilistic messages between 
the elements provides a very detailed
statistical information on the various solutions. One can expect
that this general framework will also find applications
in  neural network studies.

\subsubsection*{Acknowledgments}
The results presented here have been obtained in various collaborations, mainly
with G. Parisi and R. Zecchina, and also with A. Braunstein, S. Mertens, M.
Weigt.



\end{document}